\documentstyle[aps,floats,epsf,prb,preprint]{revtex}
\begin{document}
\draft
\tighten
\title{Convective Term and Transversely Driven Charge-Density Waves}
\author{S.N. Artemenko, S.V. Zaitsev-Zotov, V.E. Minakova}
\address{
Institute for Radioengineering and Electronics of Russian
Academy of Sciences, Mokhovaya 11, 103907 Moscow, Russia\\
}
\author{P. Monceau}
\address{
Centre de Recherches sur les Tr\'es Basses Temp\`eratures, CNRS, BP
166, 38042 Grenoble, France\\
}
\maketitle
\begin{abstract}
We derive the convective terms in the damping which determine the
structure of the moving charge-density wave (CDW), and study the
effect of a current flowing transverse to conducting chains on the
CDW dynamics along the chains. In contrast to a recent prediction
we find that the effect is orders of magnitude smaller, and that
contributions from transverse currents of electron- and hole-like
quasiparticles to the force exerted on the CDW along the chains
act in the opposite directions. We discuss recent experimental
verification of the effect and demonstrate experimentally that
geometry effects might mimic the transverse current effect.
\end{abstract}
\pacs{PACS numbers: 71.45.Lr, 72.15.-v, 72.20.Pa}

Nonequilibrium steady states of driven periodic systems with
quenched disorder have attracted significant interest. Examples
include quasi-1D charge-density waves (CDW) \cite{1}, vortex
lattices in superconductors \cite{2}, two-dimensional Wigner
crystals \cite{3}. All these systems are characterized by pinning,
so that macroscopic motion is only set up when the applied force
is larger than a threshold value.

In the particular case of quasi-1D CDW where the CDW motion is
only along the chains, many interesting effects were predicted
\cite{4,5,11} such as a dynamical phase transition between a
disordered and an ordered state and instability due to the
proliferation of phase slips. While a dynamic transition was
observed in the recent experiment on high-quality NbSe$_3$
crystals \cite{Th} between temporally ordered CDW creep to
high-velocity sliding, many of the predicted effects have not yet
been detected. The predictions were based on the equation of
motion including a phenomenological convective term in the
damping, $\gamma[\partial_t \varphi + ({\bf v} \nabla)\varphi]$,
where ${\bf v}$ stands for the velocity of the moving lattice (in
our case for the CDW velocity).

Recently Radzihovsky and Toner \cite{Radzi} predicted a dramatic
effect of transverse current on CDW motion along the conducting
chains. This effect results in a "current-effect transistor" in
which the CDW channel is turned on by a transverse quasiparticle
current. The effect, again, stemmed from a convective term
phenomenologically inserted in the equation of CDW motion, the
velocity ${\bf v}$ being assumed to be of the order of the normal
electron velocity. An experimental verification of this effect was
reported recently by Markovic {\em et al} \cite{Marcovic}.

The convective term introduced in ref.\onlinecite{Radzi} has a
different origin from that of refs.\onlinecite{4,5,11}. In the
former case such a term reflects the action of the transverse
current on the CDW, in the latter case it appeared because
Galilean invariance was implicitly assumed for damping, and the
time derivative $\partial_t$ was replaced by the convective
derivative $\partial_t + v \partial_x$. This would be valid if
damping in the laboratory frame were the same as in the frame
moving with the CDW. CDW damping is induced by scattering of
quasiparticles, thus the scattering would be the same in both
frames if the scatterers (impurities and phonons) were moving with
the CDW. Since it is not so, calculation of the damping
coefficient from the microscopic theory is needed. The explicite
calculation \cite{AV} yields neither Galilean invariance nor the
convective term. Nevertheless, terms with such a symmetry are
allowed, and terms resembling the convective one appear for a
completely different physical reason, namely, due to local
perturbations of the quasiparticle density. We discuss this
problem shortly below.

The main purpose of our paper is to elucidate the physical origin
and to derive the convective term by means of the microscopic
transport theory~\cite{AV}. We find that this term is orders of
magnitude smaller than it was proposed in ref.\onlinecite{Radzi};
furthermore, it is proportional to the {\it difference} of
transverse electron and hole currents rather than to the total
transverse current. Thus the net force acting on the CDW is
determined by the difference of quasiparticle contributions from
energies above and below the Fermi surface, like in the Hall
effect or thermopower. In principle, predictions of
ref.\onlinecite{Radzi} survive, but the effects become much
smaller. As for the experimental verification of this effect
\cite{Marcovic}, we show that a similar behavior may be caused by
quite large longitudinal current induced by the transversely
applied voltage due to the large anisotropy of CDW compounds.

Though we study here the CDW only, we expect that our conclusions
are qualitatively valid for other types of moving periodic
structures. In these systems the damping is related to
quasiparticle scattering as well, and the Galilean invariance does
not hold, so the longitudinal convective term must be proved by a
microscopic theory. As for the transverse effect in related
systems, we expect that it must also be related to the difference
between the electron and hole current rather than to the total
current, as well.

To derive the equation of motion for the CDW phase we start from the
equation for the nonequilibrium quasiclassical Green's functions in Keldysh
representation \cite{AV} which we use in discrete form with respect to chain
indices \cite{AZh}. The Green's functions
\[ \check{g}(t_1,t_2) =\left( \begin{array}{cc}
\hat{g}^R & \hat{g}^K\\
0         & \hat{g}^A
\end{array}
\right)\]
are formed by the retarded and advanced Green's functions and by $\hat{g}^K$
which is responsible for the electron distribution. Each of
$\hat{g}^J$ is a matrix with respect to the index identifying the sheets of
the Fermi surface of the quasi-one-dimensional conductor at $+p_F$ and
$-p_F$, and to the chain number. They satisfy the equation (in units with
$e=1$, $\hbar=1$)
\begin{eqnarray}
&&i v_F \partial_x\check{g}_{nm}+\sum_i t_{\perp i} [A_{nn+i}(t_1)
\check{g}_{n+i\,m}-\check{g}_{n\,m+i} A_{m+im}(t_2)]
 \nonumber \\
&&+h_n(t_1) \check{g}_{nm}-\check{g}_{nm} h_m(t_2) +
\Sigma_n \check{g}_{nm} -
\check{g}_{nm} \Sigma_m =0,
 \label{1}
\end{eqnarray}
where matrix product and the convolution with respect to time in the last
terms presenting the elastic collision integral are assumed, $h_n =
i\sigma_z\partial_t+ i\sigma_y\Delta_n -\Phi_n\sigma_z,\;
\Phi_n=\phi_n- (v_F/2)\partial_x \varphi_n,$ $\phi_n$ is the electric
potential, $\Delta_n$ and $\varphi_n$ are the CDW amplitude and phase,
$\Sigma_n = \frac{i}{2}
\nu_f\sigma_z\check{g}_{nn}\sigma_z-\frac{i}{4}
\nu_b[\sigma_x\check{g}_{nn}\sigma_x +\sigma_y\check{g}_{nn}\sigma_y]$,
$\nu_f$ and $\nu_b$ are the forward- and back-scattering rates. The term
with $t_{\perp}$ describes the interchain coupling.
$A_{nm}=\sigma_z\cos{\frac{\varphi_{n,m}}{2}}+
i\sin{\frac{\varphi_{n,m}}{2}}$, $\varphi_{n,m}=\varphi_n-\varphi_m$.

Equation for the phase can be found from the self-consistency conditions for
the order parameter \cite{AV,AZh}
\begin{equation}\label{sc}
{\rm Tr} \sigma_x \hat g^K_{nn} (t,t) =i \frac{m^*}{m}\frac{1}{\Delta}
\frac{\partial^2 \varphi_n}{\partial t^2}.
\end{equation}

We solve equation (\ref{1}) perturbatively with respect to $t_{\perp}$. The
function $g^{K}_{nn}$ we present as a sum of regular and anomalous parts,
the latter is related to nonequilibrium perturbations of the quasiparticle
distribution function. $\hat{g}^K_{nm} = \hat{g}^R_{nm}t_m -
\hat{g}^A_{nm}t_n + \hat{g}^{(a)}_{nm}$, $t_n= 1-2n_F \,(\epsilon- \mu_n)$,
$n_F$ is the Fermi distribution function, $\mu_n$ is the shift of the
chemical potential in chain $n$, and $\epsilon$ is the energy related to
$t_1-t_2$ by Fourier transformation.

In the zeroth approximation we neglect $t_{\perp}$ and find \cite{AV,AZh}
$g^{R(A)}_{nn}= g_n\sigma_z+f_ni\sigma_y$, $g^{(a)}_{nn}= 0$ with
$g_n=\epsilon_n/\xi_n$, $f_n=\tilde{\Delta}_n/\xi_n$,
$\epsilon_n=\epsilon-\Phi_n+i\nu_+g^{R(A)}_n/2$, $\nu_+=\nu_f+\nu_b$,
$\tilde{\Delta}_n=\Delta_n-i\nu_ff^{R(A)}_n/2,\;
\xi_n^{R(A)} =\pm \sqrt{\epsilon_n^2-\tilde{\Delta}_n^2} $.

In the first approximation we calculate Green's functions off-diagonal with
respect to layer indices.
\begin{equation}
\hat g_{n\,n+i}^{(a)}=\sum_i
\frac{t_{\perp,i}(t_n-t_{n+i})}{\xi_n^{R}+\xi_{n+i}^A}(A_{n\,n+i}- \hat
g_{nn}^R A_{n,n+i} g_{n+i\,n+i}^A),
\label{1t}
\end{equation}
the result for retarded (advanced) function can be obtained from (\ref{1t})
by substituting $(t_n-t_{n+i})$ by 1, and all indices $A$ for $R$ ($R$ for
$A$).

Using (\ref{1t}) we can calculate the transverse current.
$$
j_{n,n+1}(x) =\frac{i \kappa^2}{64 \pi} \int t_{\perp i} {\rm
Tr}(A_{nn+1}\hat{g}^K_{n+1\,n}- \hat{g}^K_{n\,n+1} A_{n+1n})\,d\epsilon,
$$
where $\kappa$ is the inverse Thomas-Fermi screening radius. In the
linear approximation with respect to the driving field, $E_\perp = (V_n -
V_{n+1})/d$, related to the difference of electrochemical potentials, $V_n =
\Phi_n - \mu_n$, we get for the simplest case of identical chains and
temperatures $T \ll \Delta$
\begin{eqnarray}
&&  j_{n,n+1}(x) = \sigma_{\perp}E_\perp,  \label{jtE}  \\
&& \sigma_{\perp}= \sigma_{\perp N} \frac{\nu_+}{2\nu}
\left(\cos^2{\frac{\varphi_{n,n+1}}{2}} +\frac{2T}{\Delta}
\right)e^{-\Delta/T} \sinh{\frac{\mu}{T}} , \nonumber
\end{eqnarray}
where $\nu_+=\nu_f+\nu_b$, $\nu =\nu_f + \nu_b$, $\sigma_{\perp N}$ is the
normal-state conductivity. Under typical experimental conditions the
phase in the transverse directions varies over large distances and one can
substitute $\varphi_{n,n+1}$ for zero.

If some chains are not in the Peierls state, like in NbSe$_3$, an
additional contribution to the conductivity appears, which is related to the
hopping between the chains in the metallic state and exhibits no
thermal activation.

The second order corrections in $t_\perp$ to $\hat g_{nn}$ we calculate by
inserting the first order correction (\ref{1t}) into equation (\ref{1}) for
$n=m$. After some algebra we get the component of Green's function $g_x =
{\rm Tr}\hat g_{nn}^{K}$ needed to calculate the equation for the phase
by means of eq.(\ref{sc}). In addition to standard terms describing the
elastic deformation, action of the electric field and damping of the CDW
which can be found in refs.\onlinecite{AV,AZh}, this function contains the
term which is related to perturbations of the quasiparticle distribution
function by the transverse quasiparticle flow.  This additional term reads
\begin{eqnarray}
&&  g_{x\perp}^{(a)}=\sum_i
i t_{\perp i}^2(V_{n+i}-V_{n})\sin{\varphi_{n,n+i}}
G(\varepsilon), \label{gx} \\
&& G \! = \! \frac{d t_n }{d \varepsilon}  {\rm Re}
\frac{f^R_{n+i}[\xi_{n+i}^R + (\xi_n^{R}+\xi_{n}^A+\xi_{n+i}^R)(g^R_{n}g^A_{n} -
f^R_{n}f^A_{n})]}{(\xi_n^{R}+\xi_{n}^A)(\xi_n^{R}
+\xi_{n+i}^R)(\xi_n^{A}+\xi_{n+i}^R)}. \nonumber
\end{eqnarray}
Integration of expression (\ref{gx}) over energy gives the force induced by
the transverse currents. Note that in $G(\epsilon)$ the derivative of the
Fermi distribution functions is multiplied by an odd function of energy.
This means that the contributions of electrons ($\epsilon >0$) and
holes ($\epsilon <0$) have different signs. If the chemical potential is not
shifted from the midgap position, {\em i. e.} $\mu=0$, then the electron and
hole contributions compensate each other, and the convective term
disappears. If $\mu \neq 0$ then this compensation is not complete. For
brevity we assume the condition $T \ll \Delta$ which typically holds in CDW
conductors practically at all temperatures below the fluctuation region near
the Peierls transition. Then the equation for the phase acquires the form
\begin{eqnarray}
&&
\frac{m*}{m}\frac{\partial^2 \varphi_n}{\partial t^2}  +
\gamma\frac{\partial \varphi_n}{\partial t} -
v_F^2\frac{\partial^2\varphi_n}{\partial x^2} \\ \nonumber
&& +2 \sum_{i} t_{\perp,i}^2
[1 + A (V_n - V_{n+i})] \sin{\varphi_{n,n+i}} = F_{pin}+ 2 v_F E_n,\label{pha}
\end{eqnarray}
where $A=(\Delta_n/2) \int G(\epsilon ) d\epsilon$, and $F_{pin}$ is a pinning
force. In the continuous approximation equation the last term in the
left-hand side of (\ref{pha}) reads $v_\perp^2 [\nabla_\perp^2
\varphi +2A ({\bf E}_\perp \nabla_\perp) \varphi],$
where $v_\perp^2 =(t_\perp d)^2/2$, $d$ is the interchain distance. The term
with $A$ is the convective term describing the transverse effect. Coefficients $\gamma$ and $A$ in the equation depend on details
of the energy structure of the compound, {\em e. g.} on energy dependence
of the density of states {\em etc}. Noting that the main contribution to the
integrals stems from the energies close to the gap edge, we simplify
(\ref{gx}) in two limits,
$p=(\Delta \nu^2)^{1/3}/T \ll 1$ and $p \gg 1$.  With the accuracy of a
factor of the order of unity we get
\begin{eqnarray}
&&
\gamma \sim 2\nu_b \frac{\Delta}{T} \left\{ \begin{array}{c}  \ln{1/p} \\
1/p^2 \end{array} \right\} e^{-\Delta/T} \cosh{\frac{\mu}{T}},  \nonumber \\
&& A \sim \frac{1}{\nu} \left\{ \begin{array}{c} p \\ 1/\sqrt{p}
\end{array} \right\} e^{-\Delta/T} \sinh{\frac{\mu}{T}} \quad \mbox{at}
\;
\left\{ \begin{array}{c} p \ll 1 \\ p \gg 1 \end{array} \right\} \label{ga}
\end{eqnarray}
Both the damping factor $\gamma$ and the convective term exhibit a thermally
activated behavior because they are related to excitations via the Peierls
gap. If some chains are not in the Peierls state,
non-thermally activated contributions appear in the next order
approximations in $t_\perp/\Delta$. This happens due to proximity effect
which induces non-zero density-of-states inside the gap at the chains in the
Peierls state and to CDW correlations at the chains in the metallic state.
Such contributions are of the order of $\nu_b t_\perp^2/\Delta^2$ to factor
$\gamma$, and of the order of $t_\perp^2 \mu/\Delta^4$ to coefficient $A$.
Though these contributions are small they may become dominant since
contributions (\ref{ga}) are thermally activated and may be very small.

Eq.(\ref{ga}) shows that the local value of the damping coefficient depends
on a local variation of the chemical potential, the latter being related to
a local perturbation of the quasiparticle density $\delta n / n$
\begin{equation}
\gamma = \gamma_0 (1+ a \delta n / n), \label{ga1}
\end{equation}
where $a$ is a factor of order 1. The last term can be presented in the form
of a convective term with the velocity parallel to the CDW motion
\cite{4,5,11}. We present a "quick and dirty"
qualitative derivation of such a term. Let us consider the regime of linear
screening when the shift of the chemical potential $\delta\mu$ from the
equilibrium position is small. Then we estimate $\delta n / n \sim
\delta\mu/T$ in a semiconducting CDW state, and $\delta n / n \sim
\delta\mu/E_F$ in the metallic CDW state, where $E_F$ is the Fermi energy.
$\delta \mu$ can be found from the quasineutrality condition \cite{AV}. The
local perturbation of the charge density consists of the CDW and
the quasiparticle contribution, $\rho \propto [N_c v_F \partial_x
\varphi/2 + (1- N_c)\delta\mu$], where $N_c$ is the fraction of
electrons condenced in the CDW. Then, expressing $\delta \mu$ from the
condition $\rho=0$ and inserting it to (\ref{ga1}), we find the
renormalisation of the damping term. Using the standard expression for the
CDW velocity $v=\partial_t \varphi /2k_F$ we present the damping term for
the metallic CDW in the form resembling the convective derivative
\begin{equation}\label{co}
\gamma=\gamma_0 \left( \partial_t\varphi+ b \frac{N_c}{1-N_c} v
\partial_x \varphi \right).
\end{equation}
The "convective" term here is proportional to the temperature dependent
ratio between contribution of the condensed electrons and quasiparticles,
which is small in NbSe$_3$ or {\it m}-TaS$_3$ for the upper CDW region. In
the case of semiconducting CDW the additional factor $E_F/T$ in the
"convective" term appears, and since typically $(1-N_c) \ll 1$ the
renormalisation of the damping coefficient is large. Note, however, that in
refs.\onlinecite{4,5,11} the convective term contains an averaged CDW
velocity, while in Eq.(\ref{co}) $v$ is the time and position dependent
local CDW velocity which can be treated as a constant parameter only in the
limit of large CDW velocity. Thus electric fields induced by local
perturbations of the quasiparticle density renormalize the damping similarly
to renormalization of the quasiparticle conductivity (\ref{jtE}) and of the
CDW stiffness~\cite{AV}. A detailed discussion of the effects induced by
this convective term is beyond the scope of the present paper, therefore, we
limit the discussion by a comment that some predictions of
refs.\onlinecite{4,5,11} must be revised.

Now we return to the transverse convective term. Its sign depends on the
sign of the charge of the quasiparticles which determine the single-particle
conductivity, and its temperature dependence is determined by an interplay
of a thermally activated behavior and higher order terms in the interchain
interaction. The exact value of the convective term depends on the details
of the electron spectrum of the material.  Using calculated values for the
factors in equation of motion (\ref{pha}) one can estimate the magnitude of
the transverse electric field which may affect the CDW dynamics along the
chains. In particular, following the calculations of ref.\onlinecite{5} one
gets a decrease of the threshold depinning field along the chains as a
function of the perpendicular electric field $E_\perp$ at $E_\perp \gg E_c$,
where $E_c$ is the crossover field
\begin{equation}\label{ET}
E_T(E_\perp) = E_T(0)\frac{E_\perp}{E_c} e^{-2E_\perp/E_c}, \quad
E_c=\frac{\sqrt{v_F E_T(0)}}{2A v_\perp}.
\end{equation}
For the case of a semimetallic CDW material like NbSe$_3$ when
non-thermoactivated contribution to the factor $A$ dominates, the crossover
field can be estimated as $$E_c \approx \frac{\Delta^3}{t_\perp^2 \mu}
\frac{v_F}{v_\perp}
\sqrt{E_\Delta E_T(0)}, \quad E_\Delta = \frac{\Delta^2}{v_F} \sim
10^4 \div 10^5 \mbox{V/cm}.$$
The crossover field $E_c$ is much larger than the longitudinal
threshold field $E_T(0)$ because both $E_\Delta$ and the
prefactor before the square root are large.

The experimental observation of the effect of a transverse current on the
longitudinal CDW dynamics was reported recently by Markovic {\it et
al.}~\cite{Marcovic} in NbSe$_3$. It was
found that application of a transverse electric field of the order of 1~V/cm
was enough for a substantial reduction of the longitudinal threshold field.
The discrepancy with the above estimate is thus 4-5 orders of magnitude.
Now we discuss the problems which may occur in the experimental study of the
effect in the simple geometry with two side contacts for the application of the
transverse current~\cite{Marcovic}. Let us consider a sample of width $w$
along $y$ axis
with side current contacts of length $l$ along the $x$ axis parallel to the
chain direction. Assuming a uniform current density in the contacts and
Ohmic conductivities $\sigma_\| \gg \sigma_\perp$ in $x$ and $y$ directions
($y=0$ at the contact),
respectively, we calculate the distribution of the electric potential in the
sample (see ref.\onlinecite{zz})
\begin{eqnarray}
\label{ln}
\phi \! = \! E_0 \!\!\! \int\limits_{-l/2}^{l/2} \!\! dx_1 \!\!\!
\sum_{n=-\infty}^{\infty} \!\! \ln{\frac{\sqrt{(x-x_1)^2\sigma_\perp
+(y+2nw)^2\sigma_\|}}{\sqrt{(x-x_1)^2\sigma_\perp
+(y-2nw)^2\sigma_\|}}} ,\!\!
\end{eqnarray}
where $E_0=j_\perp/(2 \pi \sigma_\perp)$, $j_\perp$ is the
transverse current density through the contact.

In the case of large conductivity anisotropy, large
components of the longitudinal current appear near the contacts.  According
to (\ref{ln}) in the region near the contacts, at distances of the order of
$l$ along the chains and at distances about $l\sqrt{\sigma_\|/\sigma_\perp}$
in the transverse direction, the component of the electric field parallel to
the chains is almost constant, $E_\| \approx E_0$.
At larger distances the electric field decays as $2E_\| l/x$ at distances
from $x \gg l, y \sqrt{\sigma_\|/\sigma_\perp}$, and as $2E_\|
lx/y^2$ at $x,l \ll y \sqrt{\sigma_\|/\sigma_\perp}$.  And,
finally, it decays rapidly at $x > w\sqrt{\sigma_\|/\sigma_\perp}$.

\begin{figure}[h]
\vskip -1.8cm \epsfxsize=10cm \centerline{\epsffile{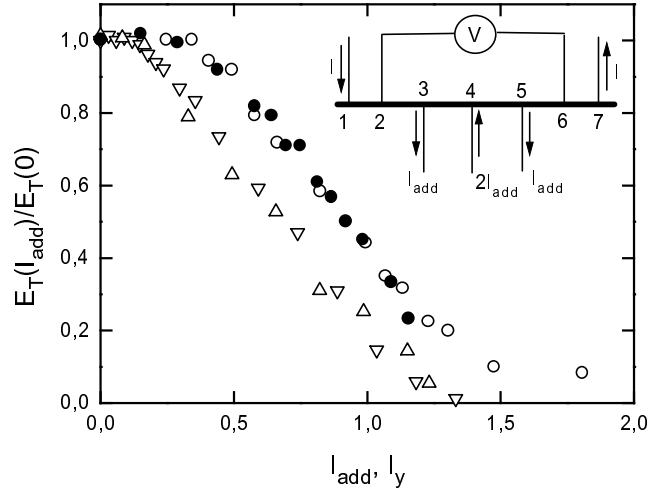}}
\vskip -5.7cm \caption{Dependence of the threshold field of
segment 2-6 on an additional {\it longitudinal} current in a
geometry shown in inset for two samples of NbSe$_3$ with contact
separations $l_{23},l_{34},l_{45},l_{56}\approx100$~$\mu$m:
$\bigtriangledown$ - $T=115$~K, $\bigtriangleup$ - $T=128$~Š.
$\circ$ and $\bullet$ - dependence $E_T(I_y)/E_T(0)$ on the
transversal current $I_y$ (data of
ref.\protect\onlinecite{Marcovic}). $I_{add}$ is plotted in units
$I_T$ for our data, and $I_y$ is multiplied by $6\times
10^{-4}$~cm$^2$/A for data of ref.\protect\onlinecite{Marcovic}.}
\label{fig2}
\end{figure}

Thus, if a transverse voltage $V$ applied to the sample is large
enough, $E_\|$ may be larger than the threshold depinning field
$E_T$. Moreover, the region where $E_\|>E_T$ increases with
increasing $V$. Note also that the potential difference applied at
different directions from the contacts along the chains is equal
to $V/2$: the potential $\phi$ decays along the $x$ axis at
distances of the order of $w\sqrt{\sigma_\|/\sigma_\perp}$ from
the contact ({\it cf.} (\ref{ln})). This voltage may easily exceed
the threshold voltage needed to drive the CDW along the chains. In
the experiments of ref.\onlinecite{Marcovic} $l=100$~$\mu$m,
$w=36$~$\mu$m, $\sigma_\perp /\sigma_\| \approx 0.04$. Then for
the characteristic value of the transverse current $\sim
200$~$\mu$A shown to suppress the longitudinal threshold field, we
estimate $E_\| \sim 0.3$~V/cm.  That is well enough for depinning
the CDW for which $E_T\approx 0.2$ V/cm even if one takes into
account the phase-slip voltage \cite{Gill} of the order of 1 mV
required for CDW breaking. So we conclude that the effect of the
transverse electric field on the CDW dynamics can be easily masked
by the CDW depinning by the longitudinal electric field component
which is quite large due to large anisotropy of the conductivity.
Moreover, when an additional {\it longitudinal} current $I_{add}$
is applied to the central segment of the sample, the dependence of
the threshold field on $I_{add}$ mimics that of the transverse
current observed in ref.\onlinecite{Marcovic} (see
Fig.~\ref{fig2}). It also shows the crossover current for the
onset of the decrease of $E_T(I_{add})$ considered in
ref.\onlinecite{Marcovic} as an argument against the possibility
that the threshold field reduction is due to current
inhomogeneities around the transverse contacts. The crossover
behavior may be caused by the phase-slip voltage in the central
segment where the currents $I$ and $I_{add}$ have the same
direction (see inset in Fig.~\ref{fig2}).

In conclusion, the calculated convective term is found to be much
smaller than it was originally suggested~\cite{Radzi}.  Our
analysis of the experimental verification of the
effect~\cite{Marcovic} shows that the observed behavior may be
caused by geometric effects due to high anisotropy of the
conductivity.  So new experimental studies are needed.

We are grateful to N. Markovic and H. S. J. Van der Zant for
detailed discussion of their work. This work has been supported by
C.N.R.S. through the twinning program 19 between C.R.T.B.T. and
IRE RAS, by the Russian Foundation for Basic Research (project
98-02-16667), and by Russian program ``Physics of Solid State
Nanostructures'' (project 97-1052).

\end{document}